# Complex Spaces In Hydrodynamics: Complex Navier-Stokes Equations

A. N. Panchenkov[1]

*Nizhniy Novgorod State Technical University, 24 Minina St., Nizhniy Novgorod, Russia 603600*

The study is devoted to the development of new effective tools and methods of analytical hydrodynamics, including problems of existence, smoothness and structure of laminar and turbulent flows. The main problem – complex Navier-Stokes equations and turbulence in complex spaces. The necessity of introducing complex spaces in hydrodynamics is determined by the mechanism of transition of a laminar flow into a turbulent flow. The author proposes a non-traditional scenario of the transition: the cause of turbulence is in destruction (cessation of existence) of a laminar flow. The article contains the mathematical rationale for the necessity of development of the theory of turbulence in the complex configurational space: the complex configurational space is the natural area of existence of turbulence. Hydrodynamic flows are regarded as flows on entropy manifolds that [flows] are supported by the two symmetries: the symmetry of conservation of general entropy and the symmetry of duality of impulse representation. The new symmetry has been introduced and studied: the forminvariance of Helmholtz matrix of impulse density. The strict foundation has been provided for the known fact of chaotic mechanics: appearance of the new structure (a turbulent flow) is a result of interaction of two entities – dissipation and vorticity.

On the deep level the phenomenology of turbulence in complex spaces is based on the transition from the mechanics of a material point to the mechanics of an oriented material point, that [transition] takes place in a current period of time.

**I**. The problem of complex spaces in hydrodynamics has attracted attention of researchers due to appearance of the entropy conceptual model, methodology and tools of description of the Nature and surrounding Reality. This problem is intimately connected with analytical hydrodynamics and theory of turbulence.

Analytical hydrodynamics and its important section - the theory of Navier-Stokes equations - are attracting attention of many researchers nowadays. In the theory of Navier-Stokes equations, starting from the work by Leray [1], the majority of researches have been devoted to the problem of "weak" (turbulent) solutions (see works [2-6]). Despite the multitude of such researches, our knowledge of the structure and properties of the solutions of Navier-Stokes equations remains rather limited. Therefore, the task of brining new ideas, concepts and methodologies into analytical hydrodynamics has become actual.

The new concept, methodology and tools for studying continuous media have been developed by A.N. Panchenkov in the book series "Entropy" [7–10].

The one of characteristic features of this series is the following: in the majority of the problems the basic geometrical objects (phase spaces, configurational spaces, entropy manifolds etc. ) are complex. And what is more, the subject of the book "Inertia" - the theory of inertia - was realized above the field of complex numbers. The entropy mechanics is defined by A.N. Panchenkov in the book "Entropy mechanics" as the mechanics of flows on entropy manifolds of complex configurational spaces and fields in complex configurational spaces.

---

[1] e–mail: entropyworld@narod.ru

The factor determining advisability and necessity of the complexification is the existence of rotor in the virtual continuous medium. Let me remind that Navier-Stokes equations were introduced into hydrodynamics for describing laminar flows of a viscous incompressible fluid; the question of validity and dependability of their application is still open. The applicability of real Navier-Stokes equations for describing turbulent flows was questioned by several investigators (see, e.g. [3]). Yet this question has become especially actual lately, due to activation of researches on theory of these equations [4], [6], [http://claymath.org]. The particular facts and materials determining necessity of transition to the complex Navier-Stokes equations in the problem of turbulence are contained in the book [8]. The author set forth the non-classical hypothesis of turbulence origin in the chapter "Turbulence" of the book "Entropy-2: Chaotic Mechanics." The origin of turbulence is believed, usually, to be related to the loss of stability of a laminar motion. I accepted the another hypothesis, that is not connected whatsoever with the loss of stability of a laminar motion. A laminar motion can not exist forever; sooner or later, it ceases to exist. A time-point of cessation of existence (destruction) of a laminar flow is, simultaneously, the point of appearance of a turbulent flow.

According to this hypothesis, the cause of turbulence is a destruction of a laminar flow. A destruction of a laminar flow (as of a certain structure) is not directly related to stability of a motion; it is not a loss of stability that is implied, but destruction of one structure (laminar flow) followed by appearance of another structure (turbulent flow). Both destruction and appearance take place in chaos, that fact determines the key role of the chaotic mechanics.

The development of the effective theory of turbulence, based on the hypothesis of destruction of a laminar flow, requires, naturally, introducing the complex basic geometrical objects and the complex Navier-Stokes equations; such introduction has been performed by me in the book [8]. It has been established there, that, regarding the problem of turbulence, the complex phase space is the natural environment for the turbulence.

**II**. The basic postulate of the entropy analysis of hydrodynamics is formulated as follows: *the conceptual model of hydrodynamics is the concretization of the entropy conceptual model* [11].

The basic geometrical object of the entropy theory is the phase space - the smooth manifold with local coordinates p and q:

$$\Omega = \{ q, p \mid \Omega = \Omega_q \times \Omega_p; \Omega_q \subset R^3; \Omega_p \subset R_3; \Omega \subset R^3 \oplus R_3 \}.$$

Here:
q – the generalized coordinate.
p – the impulse.
$R^3$ – three-dimensional real Euclidean space.
$R_3$ – conjugate three-dimensional real Euclidean space
The phase space forms:
1. The configurational space

$$\Omega_q = \{ q \mid \Omega_q \subset R^3 \}.$$

2. The impulse space

$$\Omega_p = \{ p \mid \Omega_p \subset R_3 \}.$$

The general entropy has the dual representation:

$$H_f = H_q + H_p; \{q, p\} \in \Omega, \qquad (1)$$

$H_q$ – structural entropy,
$H_p$ – impulse entropy.

The global symmetry is supported on the states of medium, located in the phase space:

$$H_f = const. \qquad (2)$$

This symmetry follows from the principle of the entropy maximum applied to the Boltzmann concept of entropy [7]

$$H_f \triangleq -\int_\Omega \rho \ln \rho \, d\Omega, \{q, p\} \in \Omega. \qquad (3)$$

This initial representation of entropy contains $\rho$ – density of the virtual continuous medium.

The first reduction of the phase space – the smooth manifold called the entropy manifold - is organized by means of the symmetry (2)

$$Э = \{q, p \mid Э \subset \Omega, H_f\}. \qquad (4)$$

The entropy manifold has the structure of the direct product:

$$Э = Э_q \times Э_p : Э_q = \{q \mid Э_q \subset Э, H_q\}; Э_p = \{p \mid Э_p \subset Э, H_p\}. \qquad (5)$$

Here:
$Э_q$ – the entropy manifold of the configurational space,
$Э_p$ – the entropy manifold of the impulse space.

The solenoidal manifold is obtained via setting the divergence on the entropy manifold

$$\sigma = divA; A = \left[\frac{\partial q}{\partial t}, \frac{\partial p}{\partial t}\right];$$
$$\mathcal{M} = \{q, p \mid \mathcal{M} \subset Э, \sigma = divA\}. \qquad (6)$$

The symmetry (2) is supported on the solenoidal manifold by the equation

$$\sigma = 0; \{q, p\} \in \mathcal{M}. \qquad (7)$$

The introduction of the potential of accelerations in the theory leads to the next reduction of the entropy manifold - the manifold of the potential of accelerations

$$\mathcal{P} = \left\{q, p \mid \ddot{I} \subset \mathcal{M}; \Theta; \xi = \begin{pmatrix} 0 & -E \\ E & 0 \end{pmatrix}\right\}. \qquad (8)$$

Here:

Θ – the potential of accelerations, $\xi = \begin{pmatrix} 0 & -E \\ E & 0 \end{pmatrix}$ – the skew-symmetric matrix of the metric (the metrtic), E – the unitary diagonal matrix.

As is well known [7], the canonical equations of the potential of accelerations are held on the manifold of the potential of accelerations

$$\frac{\partial q}{\partial t} = -\frac{\partial \Theta}{\partial p}; \frac{\partial p}{\partial t} = \frac{\partial \Theta}{\partial q}; \{q, p\} \in \mathcal{J}\!\mathcal{T}. \tag{9}$$

The reduction of the manifold of the potential of accelerations, that [reduction] contains impulse potential, is an important concept in our theory.

The submanifold of the manifold of the potential of accelerations, that [submanifold] contains the impulse potential, is called Hilbert field.

Hilbert field has the following representation:

$$\mathcal{T} = \{q, p | \mathcal{T} \subset \mathcal{J}\!\mathcal{T}; \Psi\}. \tag{10}$$

On Hilbert field

$$p = \text{grad } \Psi; \Psi = \Psi(q, t)$$

and the equation of potential of accelerations is held [7], [12]

$$\frac{\partial \Psi}{\partial t} = \Theta; \{q, p\} \subset \mathcal{T}. \tag{11}$$

**III.** The second symmetry – duality of representation of impulse – plays the fundamental role in the theory, that was set forth in the article [11]:

$$p = \begin{cases} p \in \Omega_p. \\ p(q,t); q \in \Omega_q; t \in [0,T]. \end{cases} \tag{12}$$

The first component of this dualism defines the free impulse, and the second component - the attached impulse. The attached impulse is realized in the case of the existence of a diffeomorphism:

$$T_s: \ \Omega_q \to \Omega_p. \ p(q, t) \in C^\infty(\Omega_q^T); \ \Omega_q^T = \Omega_q \times (0,T).$$

In the equation (12) T is the time of destruction of a laminar flow.

Let us introduce one more symmetry, in addition to the two aforementioned symmetries. Here exists also the symmetry of duality of representation of the generalized coordinate:

$$q = \begin{cases} q \in \Omega_q. \\ q(t); \ t \in [0,T]. \end{cases} \tag{13}$$

The first component of this dualism corresponds to the Eulerian description, and the second one – to the Lagrangian description.

The consequence of the symmetry (12) is the one more fundamental symmetry: duality of representation of the extended velocity vector

$$\left[\frac{\partial q}{\partial t}, \frac{\partial p}{\partial t}\right] = \begin{cases} \left[\frac{dq}{dt}, \frac{dp}{dt}\right]; p \in \Omega_p \\ \left[\frac{\partial q}{\partial t}, \frac{\partial p}{\partial t}\right]; p \in p(q,t); \frac{dq}{dt} = \frac{\partial q}{\partial t}; \frac{dp}{dt} \neq \frac{\partial p}{\partial t}. \end{cases} \quad (14)$$

**IV.** The following data on the phenomenology of hydrodynamics are needed for further discourse:
  *1. The two entities exist in the continuous medium of viscous incompressible fluid:*
    *1). Hydrodynamic fields.*
    *2). Hydrodynamic flows.*
  *2. Hydrodynamic flows are located on the entropy manifolds.*
  *3. There are two types of hydrodynamic fields:*
    *3). Vorticity fields.*
    *4). Potential fields.*
  *4. There are two kinds of hydrodynamic flows:*
    *5). Laminar flows.*
    *6). Turbulent flows.*
  *5. Turbulence is a kind of chaos.*
  *6. Laminar flows are more smooth (regular) and they are realized on the real geometrical objects – entropy manifolds.*

Taking into account the Helmholtz expansion theorem [13] and the work [11], let us accept the following representation of a laminar flow on the entropy manifold:

$$\dot{q} \triangleq u \, ; u \triangleq p + \omega \, ; \, \text{div } u = 0 \, ; q \in \Im_q \, ; t = [0,T]$$
$$\omega = \omega(q,t) \, ; q \in \Im_q \, ; \omega \triangleq \text{rot } B \, ; B \in C^\infty(\Omega_q^T).$$
$$p = \begin{cases} p \in \Omega_p \\ p(q,t) \, ; q \in \Im_q \, ; t \in [0,T]. \end{cases} \quad (15)$$
$$p = p(q,t): p = \text{grad } \Psi \, ; \Psi = \Psi(q,t); \Psi \in C^\infty(\Omega_q^T) \, ;$$
$$T_\Im: \Im_q \to \Im_p.$$

The key role in a laminar flow (15) belongs to the hypothesis of independence, introduced by the authors of the work [11]: *vorticity does not depend on the impulse*

$$\omega = \omega(q,t) \, ; \{q, p, t\} \in \Omega_T \, ; \Omega_T = \Omega \times (0,T).$$

**V.** Having in mind the book [8], let me introduce Helmholtz matrix into the discourse:

$$\Lambda = \chi + \mu + \tilde{\Omega}, \quad (16)$$

$$\chi = \begin{Vmatrix} \chi_1 & 0 & 0 \\ 0 & \chi_2 & 0 \\ 0 & 0 & \chi_3 \end{Vmatrix}; \quad \mu = \begin{Vmatrix} 0 & \mu_1 & \mu_2 \\ \mu_1 & 0 & \mu_3 \\ \mu_2 & \mu_3 & 0 \end{Vmatrix}; \quad \tilde{\Omega} = \begin{Vmatrix} 0 & -\omega_3 & \omega_2 \\ \omega_3 & 0 & -\omega_1 \\ -\omega_2 & \omega_1 & 0 \end{Vmatrix}.$$

Here: $\Lambda$ – matrix of the impulse density; $\chi$ – expansion matrix; $\mu$ – shifting matrix; $\tilde{\Omega}$ – rotor matrix.

In terms of Helmholtz matrix, the equation of a laminar flow (15) looks like the following:

$$\dot{q} \triangleq \Lambda q \ ; \ q \in \Im_q \ ; \ t \in [0,T]. \qquad (17)$$

Let me introduce now the basic hypothesis: *a laminar flow* (17) *may be represented normally on the real entropy manifold*:

$$\dot{q} \triangleq \tilde{\Lambda} q; \ q \in \Im_q \ ; \ t \in [0,T]. \qquad (18)$$
$$\tilde{\Lambda} = \operatorname{Re} \tilde{\Lambda} \ ; \ \tilde{\Lambda} = \operatorname{diag} \{\lambda_1, \lambda_2, \lambda_3\}.$$

Whence the characterization of a laminar flow follows: the diagonal matrix $\tilde{\Lambda}$ is imposed above the field of real numbers.

Let me remind, that the matrix $\tilde{\Lambda}$ has dual representation

$$\tilde{\Lambda} = \begin{cases} \tilde{\Lambda}(q,t), & q \in \Im_q. \\ \tilde{\Lambda}(t). & \end{cases} \qquad (19)$$

In turn, matrices $\{\chi, \mu, \tilde{\Omega}\}$ have also dual representation, resembling (19).

In other terms for a laminar flow, the generalized coordinate - vector-function $q = q(t)$ on the segment $[0, T]$ - is imposed above the field of real numbers.

**VI.** The accepted scenario of transition of a laminar flow into a turbulent flow is following: the cause of appearance of a turbulent flow is cessation of existence (destruction) of a laminar flow. At the time-point T one structure (a laminar flow) undergoes destruction, at the subsequent time-points, that are close to T, the new structure (a turbulent flow) arises.

In the chaotic mechanics the entropy manifold, on which an event (destruction or appearance of a structure) takes place, is called extremal boundary layer (EBL). The theory of EBL is given in the book [8].

The fundamental property of EBL is its small length along the scale of real time $\sigma J$. Therefore, in order to keep definiteness, I shall study the event of transition of a laminar flow into a turbulent flow on the small segment $[-\sqrt{\varepsilon} \div \sqrt{\varepsilon}]: (T-t) \in [-\sqrt{\varepsilon} \div \sqrt{\varepsilon}]$. Let me note, that an event and accompanying chaos in EBL have adequate description in terms of the known theory of ultimate correctness [14]. In fact, EBL is the manifold of the local ultimate incorrectness.

**Definition 1**. *If* t *is current time and* T – *time-point of destruction of a laminar flow, then laminar flow is a smooth vector-function* {q=q(t'), t'=T–t} *on the segment* [ε÷T] – *the generalized coordinate determined above the field of real numbers.*

**Definition 2.** *If* t *is current time and* T *is time-point of destruction of a laminar flow and* $T_1$ *is character time* ($T_1 > T$), *then turbulent flow is a vector-function* {q = (t'), t' = T– t} *on the segment* [–ε ÷ (T – $T_1$)] - *the generalized coordinate determined above the field of complex numbers.*

**Definition 3.** *The transition of a laminar flow into a turbulent flow is the transformation of the smooth real vector-function* {q=q(t'); q=Req; t'=T–t} *on the segment* [ε÷T] *(the generalized coordinate) - into the complex vector-function* {q = q(t'): $I_m q \neq 0$} *on the segment* [–ε ÷(T–$T_1$)] *at* $T_1$>T.

In EBL the transition of the laminar boundary layer is described by the non-smooth vector-function q(t') on the segment $[-\sqrt{\varepsilon} \div \sqrt{\varepsilon}]$.
In turn, EBL will look like the following:

$$\Im_q^- = \{q \mid q = q(t'), t' \in [-\sqrt{\varepsilon} \div \sqrt{\varepsilon}]; \|\dot{q}\| \gg \|q\|, q \in \mathbb{R}^3; \forall t' \in [\sqrt{\varepsilon} \div \varepsilon]; q \in \mathbb{C}^3; \forall t' \in [-\varepsilon \div -\sqrt{\varepsilon}]\}.$$

(20)

Let us introduce the quite clear hypothesis of normality of a flow in EBL; under conditions of applicability of this hypothesis the equation of a normal flow will be:

$$\dot{q} = \tilde{\Lambda} q; q \in \Im_q^-; t' \in [-\sqrt{\varepsilon} \div \sqrt{\varepsilon}]; t' = T - t. \qquad (21)$$

The solution of this equation looks like the following:

$$q = q_0 e^{\phi}; \phi \triangleq \int^{t'} \tilde{\Lambda} dt; t' \in [-\sqrt{\varepsilon} \div \sqrt{\varepsilon}]; t' = T - t. \qquad (22)$$

A transition of a laminar flow into a turbulent flow exists if there exists the non-smooth vector-function on the segment $[-\sqrt{\varepsilon} \div \sqrt{\varepsilon}]$; t'∈ $[-\sqrt{\varepsilon} \div \sqrt{\varepsilon}]$; t' = T− t that has the dual representation:

$$\phi = \begin{cases} \text{Re } \phi \;; \; \forall t' > 0. \\ \text{Re } \phi + i \text{Im } \phi \;; \; \text{Im } \phi \neq 0 \;; \; \forall t' < 0. \end{cases} \qquad (23)$$

There is also another duality in the chaotic mechanics [8]:

$$\tilde{\Lambda} = \begin{cases} \tilde{\Lambda}_v \;; \; \forall t' > 0. \\ \tilde{\Lambda}_v + i \tilde{\Lambda}_\omega \;; \; \forall t' < 0. \end{cases} \qquad (24)$$

This duality is more convenient in some cases, as it reduces the problem of turbulence appearance to the problem of eigenvalues. The equation of a flow in EBL has the dual representation:

$$\dot{q} = \begin{cases} \Lambda q \;; q \in \Im_q^-. \\ \tilde{\Lambda} q. \end{cases} \qquad (25)$$

The next equation follows from that duality:

$$\Lambda q = \tilde{\Lambda} q \;; \qquad \Lambda = \chi + \mu + \tilde{\Omega} \;; \; q \in \Im_q^-. \qquad (26)$$

As is known, the diagonal matrix $\tilde{\Lambda}$ in this equation is the matrix of eigenvalues of the Helmholtz matrix. We come here to the known fact [8]:
1. $\tilde{\Lambda} = \Lambda_v$ - laminar flow
2. $\tilde{\Lambda} = \Lambda_v + i \Lambda_\omega$ - turbulent flow.

**VII**. The structure of a flow in EBL is established on the basis of analysis of elementary functions. Some elementary functions do not exist for negative values of argument above the field of real numbers. There are two character functions among them:

1. $\tau = \sqrt{t'}$ - the first character function;
2. $\ln t'$ – the second character function.

The character functions for negative values of argument are represented as follows:
1. $\tau = i\sqrt{|t'|}$;   $t' < 0$.
2. $\ln t' = \pi i + \ln|t'|$; $t' < 0$.

Being functions on the segment $[-\sqrt{\varepsilon} \div \sqrt{\varepsilon}]$, the character functions are non-smooth functions having dual representations:

$$1.\ \tau = \begin{cases} \sqrt{t'}\ ; \forall\ t' > 0. \\ i\sqrt{|t'|}\ ; \forall\ t' < 0. \end{cases} \quad (27)$$

$$2.\ \ln t' = \begin{cases} \ln t'\ ; \forall\ t' > 0. \\ \pi i + \ln|t'|\ ; \forall\ t' < 0. \end{cases}$$

This information about the character elementary functions makes ground for choosing the new argument (parameter of parametrization) in EBL $\tau = \sqrt{(T-t)}$.

In this case the matrix of impulse density has the dual representation:

$$\Lambda = \begin{cases} \Lambda(q,\tau),\ q \in \Im_q. \\ \Lambda(\tau). \end{cases} \quad (28)$$

Let me now endow the matrix of impulse density with the symmetry – forminvariance. The forminvariance of the matrix of impulse density means that that the matrix-function $\Lambda$ has the same outlook (structure) for both components of time duality:

$$\tau = \begin{cases} \sqrt{T-t}\ ; \forall (T-t) > 0. \\ i\sqrt{|T-t|}\ ; \forall (T-t) < 0. \end{cases}$$

**Theorem 1**. *If there exist the symmetry – forminvariance of Helmholtz matrix – then Navier-Stokes equations:*

$$\frac{Du}{Dt} = -\operatorname{grad} \Pi + \nu \Delta u + f;\ q \in \Im_q \quad (29)$$

*possess symmetry – keep the structure for both components of time duality:*

$$\tau = \begin{cases} \sqrt{T-t}\ ; \forall (T-t) > 0. \\ i\sqrt{|T-t|}\ ; \forall (T-t) < 0. \end{cases} \quad (30)$$

*The discrimination of types of flows is made via concretization of the time $\tau$ and setting the structure of the entropy manifold of the configurational space.*

*As a result, the equations of flows look like the following:*
*1. Laminar flow:*

$$\frac{dq}{d\tau} = \Lambda_\tau q;\ \Lambda_\tau = -2\tau \Lambda;\ q \in \Im_q,$$

$$\left(\frac{1}{2\tau}\frac{d\Lambda}{d\tau} - \Lambda^2\right)q = -\text{grad }\Pi + \nu\Delta u + f;\ q \in \Im_q, \quad (31)$$
$$\text{div }\Lambda q = 0;\ \Lambda = \chi + \mu + \tilde{\Omega};$$

$$\Im_q = \{q|\ \Im_q \subset \Omega_q;\ \Omega_q \subset R^3\};\ \tau = \sqrt{(T-t)}\ ;\ t \in (0 \div T).$$

$$\chi = \begin{Vmatrix} \chi_1 & 0 & 0 \\ 0 & \chi_2 & 0 \\ 0 & 0 & \chi_3 \end{Vmatrix};\ \mu = \begin{Vmatrix} 0 & \mu_1 & \mu_2 \\ \mu_1 & 0 & \mu_3 \\ \mu_2 & \mu_3 & 0 \end{Vmatrix};\ \tilde{\Omega} = \begin{Vmatrix} 0 & -\omega_3 & \omega_2 \\ \omega_3 & 0 & -\omega_1 \\ -\omega_2 & \omega_1 & 0 \end{Vmatrix}.$$

*2. Turbulent flow:*

$$\frac{dq}{d\tau} = \Lambda_\tau q;\ \Lambda_\tau = -2\tau\Lambda;\ q \in \Im_q,$$
$$\left(\frac{1}{2\tau}\frac{d\Lambda}{dt} - \Lambda^2\right)q = -\text{qrad }\Pi + \nu\Delta u + f;\ q \in \Im_q, \quad (32)$$
$$\text{div }\Lambda q = 0;\ \Lambda = \chi + \mu + \tilde{\Omega};$$
$$\Im_q = \{q|\ \Im_q \subset \Omega_q;\ \Omega_q \subset \mathbb{C}^3\};\ \tau = i\sqrt{|T-t|}\ ;\ \forall\ (T-t) < 0,$$

$$\chi = \begin{Vmatrix} \chi_1 & 0 & 0 \\ 0 & \chi_2 & 0 \\ 0 & 0 & \chi_3 \end{Vmatrix};\ \mu = \begin{Vmatrix} 0 & \mu_1 & \mu_2 \\ \mu_1 & 0 & \mu_3 \\ \mu_2 & \mu_3 & 0 \end{Vmatrix};\ \tilde{\Omega} = \begin{Vmatrix} 0 & -\omega_3 & \omega_2 \\ \omega_3 & 0 & -\omega_1 \\ -\omega_2 & \omega_1 & 0 \end{Vmatrix}.$$

**The proof**. In the case of Helmholtz matrix of impulse density
$$\Lambda = \chi + \mu + \tilde{\Omega},$$
possessing the forminvariance regarding the time $\tau$, the same symmetry is intrinsic to the equation of a flow
$$\dot{q} = \Lambda q\ ;\ q \in \Im_q. \quad (33)$$

In turn, the equation of velocity flow is:

$$\ddot{q} = (\dot{\Lambda} + \Lambda^2)\ q\ ;\ q \in \Im_q. \quad (34)$$

It is evident, that this equation possesses the forminvariance regarding the time $\tau$. As $\dot{q} = u;\ \ddot{q} = \dfrac{Du}{Dt}$, Navier-Stokes equation and the equation (34) lead to the dualism

$$\ddot{q} = \begin{cases} (\dot{\Lambda} + \Lambda^2)\ q, & q \in \Im_q. \\ -\text{grad }\Pi + \nu\Delta\Lambda q + f. \end{cases} \quad (35)$$

This dualism leads, in turn, to the new representation of Navier-Stokes equations:

$$(\dot{\Lambda} + \Lambda^2)q = -\text{grad }\Pi + \nu\Delta u + f; \quad (36)$$
$$\Lambda = \chi + \mu + \tilde{\Omega},\ q \in \Im_q.$$

As $\dot{\Lambda} = \left(-\dfrac{1}{2\tau}\dfrac{d\Lambda}{d\tau}\right)$, in the case of the forminvariance of the matrix of impulse density the equation (36) possesses the symmetry of the structure.

Let us accept now the following characterization of flows:

1. A laminar flow

$$\tau = \sqrt{T-t}\ ;\ \forall (T-t) > 0;$$
$$\Im_q = \{q|\ \Im_q \subset \Omega_q;\ \Omega_q \subset R^3\}. \tag{37}$$

2. A turbulent flow:

$$\tau = i\sqrt{|T-t|}\ ;\ \forall (T-t) < 0\ ;$$
$$\Im_q = \{q|\ \Im_q \subset \Omega_q;\ \Omega_q \subset \mathbb{C}^3\}. \tag{38}$$

Here: $R^3$ – three-dimensional Euclidean space;

$\mathbb{C}^3$ – complex three-dimensional real Euclidean space.

Introducing these characterizations into the equation (36) and adding the equations of continuity and flow (33), we come to the equations of laminar and turbulent flows of the theorem. ∎

**VIII.** The equation of the normal flow of duality may be written as:

$$\frac{d\ln q}{d\tau} = \tilde{\Lambda}_\tau\ ;\ \tilde{\Lambda}_\tau = -2\tau\tilde{\Lambda}\ ;\ q \in \Im_q. \tag{39}$$

Now, in accordance with the book [7], we get the equation of structural entropy from the equation (39):

$$\frac{dH_q}{d\tau} = \sigma_\tau;\ H_q = (\ln q | E)_{R3};\ \sigma_\tau = Sp\ \tilde{\Lambda}_\tau\ ;\ q \in \Im_q. \tag{40}$$

Here $H_q$ - structural entropy, $\sigma_\tau = \sigma_\tau(\tau)$ – divergent invariant.

The equation (40) forms the manifold:

$$V_q = \{q|\ V_q \subset \Im_q;\ \Pi q = f_\tau;\ \Pi q = q_1 \cdot q_2 \cdot q_3\}\ . \tag{41}$$

On the manifold $V_q$, $f_\tau = f_\tau(\tau)$ - the smooth function on the segment $[\sqrt{\varepsilon} \div \sqrt{T}\,]$, $\forall (\,T - t\,) > 0$

$$f_\tau = \Pi_o + e^{\int_{}^{\tau} \sigma_\tau\, d\tau}\ .$$

Theorem 2. *On the real manifold*

$$V_q = \{q|\ V_q \subset \Im_q;\ \Pi q = f_\tau;\ \Pi q = q_1 \cdot q_2 \cdot q_3\}$$

*the equation*

$$\Lambda_\tau q = \tilde{\Lambda}_\tau q \tag{42}$$

*for the quadratic matrix [3x3]:*

$$\Lambda_\tau = \Lambda_\tau^0 + \Lambda_\tau^\nu + \Lambda_\tau^\omega$$

$$\Lambda_\tau^0 = \begin{Vmatrix} \lambda_{\tau1}^0 & 0 & 0 \\ 0 & \lambda_{\tau2}^0 & 0 \\ 0 & 0 & \lambda_{\tau2}^0 \end{Vmatrix}; \quad \Lambda_\tau^\nu = \begin{Vmatrix} 0 & \nu_{12} & \nu_{13} \\ \nu_{12} & 0 & \nu_{23} \\ \nu_{13} & \nu_{23} & 0 \end{Vmatrix}; \quad \Lambda_\tau^\omega = \begin{Vmatrix} 0 & \omega_{12} & \omega_{13} \\ -\omega_{12} & 0 & \omega_{23} \\ -\omega_{13} & -\omega_{23} & 0 \end{Vmatrix}$$

*and*

$$\tilde{\Lambda}_\tau = \text{diag}\{\lambda_\tau^1, \lambda_\tau^2, \lambda_\tau^3\}$$

*is transformed into the system of the algebraic equations:*

$$\lambda_{\tau1}^0 + (\nu_{12} + \omega_{12})\frac{q_2^2 q_3}{\Pi q} + (\nu_{13} + \omega_{13})\frac{q_2 q_3^2}{\Pi q} = \lambda_\tau^1. \tag{43}$$

$$\lambda_{\tau2}^0 + (\nu_{12} - \omega_{12})\frac{q_1^2 q_3}{\Pi q} + (\nu_{23} + \omega_{23})\frac{q_1 q_3^2}{\Pi q} = \lambda_\tau^2.$$

$$\lambda_{\tau3}^0 + (\nu_{13} - \omega_{13})\frac{q_1^2 q_3}{\Pi q} + (\nu_{23} - \omega_{23})\frac{q_1 q_3^2}{\Pi q} = \lambda_\tau^3.$$

*This system may be also represented in another form:*

$$a_1 q_2^2 + b_1 q_2 + c_1 = 0;$$
$$a_2 q_3^2 + b_2 q_3 + c_2 = 0;$$
$$a_3 q_1^2 + b_3 q_1 + c_3 = 0$$

$$a_1 = (\nu_{12} + \omega_{12})\frac{q_3}{\Pi q}; b_1 = (\nu_{13} + \omega_{13})\frac{q_3^2}{\Pi q}; c_1 = \lambda_{\tau1}^0 - \lambda_\tau^1$$

$$a_2 = (\nu_{23} + \omega_{23})\frac{q_1}{\Pi q}; b_2 = (\nu_{12} - \omega_{12})\frac{q_1^2}{\Pi q}; c_2 = \lambda_{\tau2}^0 - \lambda_\tau^2 \tag{44}$$

$$a_3 = (\nu_{13} - \omega_{13})\frac{q_2}{\Pi q}; b_3 = (\nu_{23} - \omega_{23})\frac{q_2^2}{\Pi q}; c_1 = \lambda_{\tau2}^0 - \lambda_\tau^1$$

*In turn, the system (44) is transformed into the system of nonlinear algebraic equations*

$$q_1 = \frac{-b_3 \pm \sqrt{b_3^2 - 4a_3 c_3}}{2a_3}; \quad q_2 = \frac{-b_1 \pm \sqrt{b_1^2 - 4a_1 c_1}}{2a_1}; \quad q_3 = \frac{-b_2 \pm \sqrt{b_2^2 - 4a_2 c_2}}{2a_2}. \tag{45}$$

*which allows, if granted the accessory condition of uniqueness of representation that [condition] is realized by means of the coupling equations*

$$b_1^2 = 4a_1 c_1; \quad b_2^2 = 4a_2 c_2; \quad b_3^2 = 4a_3 c_3. \tag{46},$$

*the solution*:

$$q_1 = -\frac{1}{2}\frac{(v_{23}-\omega_{23})}{(v_{13}-\omega_{13})}q_2 \ ; q_2 = -\frac{1}{2}\frac{(v_{13}+\omega_{13})}{(v_{12}+\omega_{12})}q_3 \ ; q_3 = -\frac{1}{2}\frac{(v_{12}-\omega_{12})}{(v_{23}+\omega_{23})}q_1 \qquad (47),$$

*possessing the symmetry*

$$\left(\frac{v_{23}-\omega_{23}}{v_{13}-\omega_{13}}\right)\cdot\left(\frac{v_{13}+\omega_{13}}{v_{12}+\omega_{13}}\right)\cdot\left(\frac{v_{12}-\omega_{12}}{v_{23}+\omega_{23}}\right) = -8 \qquad (48)$$

**Lemma 1**. *The symmetry, determined by the equation of the Theorem 2*

$$\left(\frac{v_{23}-\omega_{23}}{v_{13}-\omega_{13}}\right)\cdot\left(\frac{v_{13}+\omega_{13}}{v_{12}+\omega_{13}}\right)\cdot\left(\frac{v_{12}-\omega_{12}}{v_{23}+\omega_{23}}\right) = -8$$

*exists at* $v_{kj} \neq 0$, $\omega_{kj} \neq 0$, $\kappa = 1,2$ ; $j = 2,3 : \kappa \neq j$.

***Discussion***. The fundamental result of the Lemma 1, that [lemma] is associated with the separation theorem from the article [11], is of great importance for the theory of Navier-Stokes equations and turbulence. Its semantics is following: in the local zone of destruction (EBL) a laminar flow contains potential and vorticity components; an appearance of the new structure – a turbulent flow – is a result of an interaction of the two entities – dissipation and vorticity.

**IX.** For the normal flow in EBL (21) the equations of hydrodynamics may be written as follows:

$$\begin{aligned}&\dot{q} = \tilde{\Lambda}q; q \in \Im_q^- : t' \in [-\sqrt{\varepsilon} \div \sqrt{\varepsilon}]; t' = T-t.\\ &(\dot{\tilde{\Lambda}} + \tilde{\Lambda}^2)q = \varphi; \varphi = -\text{grad}\,\Pi + v\Delta\tilde{\Lambda}q + f;\\ &\text{div}\tilde{\Lambda}q = 0; \tilde{\Lambda} = \text{diag}\{\lambda_1, \lambda_2, \lambda_3\}.\end{aligned} \qquad (49)$$

The two cases are of the most interest in the problem of the transition:

$$1.\ |\dot{\lambda}_j| \sim |\lambda_j^2|; \qquad 2.\ |\dot{\lambda}_j| \gg |\lambda_j^2|.$$

The first case was studied in details in the book [8]; I am commencing here the study of the second case. In this case, the following approximate equations are held:

$$\dot{\tilde{\Lambda}}q = -\text{qrad}\,\Pi + v\Delta\tilde{\Lambda}q + f; q \in \Im_q^-. \qquad (50)$$

The characteristic solution, that satisfies the inequation 2., is the following one:

$$\tilde{\Lambda} = \tilde{\Lambda}_0 \ln \tau; \tau = \sqrt{T-t} \qquad (51)$$

The generalized coordinate, that is determined by the above solution, is defined by the form:

$$q = q_0 e^{-\tilde{\Lambda}_0 \tau^2 (\ln \tau - 1/2)} \qquad (52)$$

Let me note an interesting detail: the solution (52) satisfies the initial data, set at t = T. It is the sequitur from the symmetry of EBL – the local invariance. The bounces, arising at traversing the core of EBL, have the following values:

$$1.\ [q] = 0. \qquad 2.\ [\dot{q}] = \frac{q_0 \tilde{\Lambda}_0 \pi}{2} i.$$

So, the studied example shows that in a case of change of a flow structure the generalized coordinate remains continuous, whereas its derivative has a bounce.

The dualism, describing laminar and turbulent branches of a flow, follows from the equation (52):

$$q = \begin{cases} q_0 e^{-\tilde{\Lambda}_0(T-t)(\ln\sqrt{T-t}-1/2)}; & \forall (T-t) > 0 \\ q_0 e^{\tilde{\Lambda}_0|T-t|(\ln\sqrt{|T-t|}-1/2+\frac{\pi}{2}i)}; & \forall (T-t) < 0. \end{cases}$$

Please, note that the solution (47) is more general: it describes the representative set of attractors of a laminar flow, that [attractors] generate different kinds of turbulent flows.

In conclusion, let me cite the following sentence: on the deep level the phenomenology of turbulence in complex spaces is based on the transition from the mechanics of a material point to the mechanics of an oriented material point, that [transition] takes place in a current period of time [10].

**References**.

[1]. Leray J., 1934. Essai sur le mouvemtnt d'un liquide visquenx emplissant l' espace? Acta Matematica 63, 193–248.
[2]. Temam R., 1979. Navier Stokes Equations, Theory and Numerical Analysis. North – Holland.
[3]. Ladyzhenskaya O., 1969. The Mathematical Theory of Viscous Incompressible Flows. Cordon and Breach.
[4]. Ladyzhenskaya O., 2003. The sixth problem of the millennium: Navier-Stokes equations, existence and smoothness // Uspekhi Matematicheskikh Nauk. v. 58. №2. p 45–78.
[5]. Zvyagin, V.G. and Dmitrienko, V.T. 2004 Approximating-Topological Approach for Hydrodynamics Problems: Navier-Stokes System. Editorial "URSS", Moscow. (in Russian)
[6]. Constantin P., 2001. Some open problems and research directions in mathematical study of fluid dynamics, in Mathematics Unlimited: 2001 and Beyond, eds. Bjorn Engquist and Wilfried Schmid Springer–Verlag, Berlin.
[7]. Panchenkov, A.N. 1999. Entropy. Intelservis, Nizhniy Novgorod. (in Russian)
[8]. Panchenkov, A.N. 2002. Entropy-2: Chaotic Mechanics. Intelservis, Nizhniy Novgorod. (in Russian)
[9]. Panchenkov, A.N. 2004. Inertia. GUP "MPIK", Yoshkar-Ola. (in Russian)
[10]. Panchenkov, A.N. 2005. Entropy Mechanics. GUP "MPIK", Yoshkar-Ola. (in Russian)
[11]. Panchenkov A.N., and Matveev K.I., 2005. General–Entropy Analysis of Navier–Stokes Equations. Far East Journal of Applied Mathematics, 21(1), pp.17-30.
[12]. Panchenkov A.N., 1975. The Theory of Potential of Accelerations. Nauka, Novosibirsk. (in Russian)
[13]. Helmholtz H., 1858. Uber Integrale hidrodinamischen Glechungtn weiche den Wirbelbewegun–gen entsprechen.I.rein.angew.Vath. v. 55s. 25–55.
[14]. Panchenkov A.N., 1976. The fundamentals of the Theory of the Ultimate Correctness. Nauka, Moscow. (in Russian)